\documentclass[12pt]{article}
\usepackage{fancyhdr}
\usepackage{url}
\usepackage{cite}
 {\fancyhead{}
\fancyfoot[c]{\small{\rule{17.5cm}{1pt}}}}

\begin{document}
\title{\vspace{-2.5cm}
\begin{center}
\textbf{\small{WORKSHOP REPORT}}\\\vspace{-0.5cm} \rule{17.5cm}{1pt}
\end{center}
\vspace{1cm}\textbf{Report on the SIGIR 2019 Workshop on eCommerce (ECOM19)}}

\author{
       Jon Degenhardt \\
       eBay inc., USA\\
       \and
       Surya Kallumadi \\
       The Home Depot, USA\\
       \and
       Utkarsh Porwal \\
       eBay inc., USA\\
       \and
       Andrew Trotman \\
       University of Otago, New Zealand\\
       \date{}
}

\maketitle \thispagestyle{fancy} \abstract{ The SIGIR 2019 Workshop on eCommerce (ECOM19), was a full day workshop that took place on Thursday, July 25, 2019 in Paris, France. The purpose of the workshop was to serve as a platform for publication and discussion of Information Retrieval and NLP research and their applications in the domain of eCommerce. The workshop program was designed to bring together practitioners and researchers from academia and industry to discuss the challenges and approaches to product search and recommendation in the eCommerce domain. A second goal was to run a data challence on real-world eCommerce data. The workshop drew contributions from both industry as well as academia, in total the workshop received 38 submissions, and accepted 24 (63\%). There were two keynotes by invited speakers, a poster session where all the accepted submissions were presented, a panel discussion, and three short talks by invited speakers.}

\section{Introduction}
eCommerce Information Retrieval has received little attention in
the academic literature, yet it is an essential component of some
of the largest web sites (such as eBay, Amazon, Airbnb, Alibaba,
Taobao, Target, Facebook, and others). SIGIR has for several years
seen sponsorship from these kinds of organisations, who clearly
value the importance of research into Information Retrieval. The SIGIR 2019 Workshop on eCommerce (ECOM19)\footnote{Workshop website: https://sigir-ecom.github.io/ecom2019/} brought together researchers and practitioners of eCommerce IR to discuss topics unique to it. Our primary motivation as organizers of this workshop was to create a community and act as a forum to discuss interesting research challenges in the eCommerce domain.

Search, ranking, and recommendation have applications ranging
from traditional web search to document databases to vertical search systems. This workshop explored approaches for search
and recommendations of products. Although the task is the same
as web-page search (fulfill a user’s information need), the way in
which this is achieved is very much different. On product sites (such
as eBay, Flipkart, Amazon, and Alibaba), the traditional web-page
ranking features are either not present or are present in a different
form. The entities that need to be discovered (the information that
fulfills the need) might be unstructured, associated with structure,
semi-structured, or have facets such as: price, ratings, title, description, seller location, and so on.

Domains with such facets raise interesting research challenges
such as a) ranking functions that take into account
the tradeoffs across various facets with respect to the input query
b) recommendations based on entity similarity c) recommendations
based on user location (e.g. shipping cost), and so on. In the case of
eCommerce IR these challenges require inherent understanding of
product attributes, user behavior, and the query context. Product
sites are characterised by the presence of a dynamic inventory
with a high rate of change and turnover, and a long tail of query
distribution.

Outside of search but still within Information Retrieval, the same
feature in different domains can have radically different meaning.
For example, in email filtering the presence of “Ray-Ban” along
with a price is a strong indication of spam, but within an auction
setting this likely indicates a valid product for sale. Another example is natural language translation; company names, product names,
and even product descriptions do not translate well with existing
tools. Similar problems exist with knowledge graphs that are not
customised to match the product domain.

For this workshop we invited written submissions that included the following topics:
\begin{itemize}
\item Machine learning techniques such as online learning and deep learning for eCommerce applications
\item Semantic representation for users, products and services \& Semantic understanding of queries
\item Structured data and faceted search, for example, converting unstructured data to its structured form
\item The use of domain specific facets in search and other IR tasks, and how those facets are chosen
\item Query intent, suggestion, and auto-completion
\item Temporal dynamics for Search and Recommendation
\item Models for relevance and ranking for multi-faceted entities
\item Recall-oriented search for eCommerce including deterministic sorting of results lists (e.g. price low to high)
\item Click models for eCommerce domain
\item Session aware, and session oriented search and recommendation
\item Construction and use of knowledge graphs, and ontologies for search and recommendations
\item Personalization \& contextualization, and the use of personal facets such as age, gender, location etc.
\item Indexing and search in a rapidly changing environment (for example, an auction site)
\item Efficiency and scalability
\item Diversity in product search and recommendations
\item Strategies for resolving extremely low (or no) recall queries
\item The use of external features such as reviews and ratings in ranking
\item User interfaces (mobile, desktop, voice, etc.) and personalization
\item Reviews and sentiment analysis
\item The use of social signals in ranking and beyond
\item Performance metrics and evaluation; tying lab metrics to business metrics and KPIs
\item The balance between business requirements and user requirements (revenue vs relevance)
\item Trust and security
\item Live experimentation
\item Questions and answering, chat bots for eCommerce
\item Cross-Lingual search and machine translation
\item Fashion eCommerce
\item Conversational commerce: shopping using voice assistants such as Amazon Alexa and Google Now
\item Resources and data sets
\end{itemize}

All submissions were reviewed single blind and each submission was reviewed by at least 3 reviewers.

At the workshop we also ran a data challenge.  eBay provided a set of about 900,000 real world product titles, 150 queries, and assessments.  The task was to identify which items to show when using non-relevance sorts (such as price low to high). Users of eCommerce search applications often sort by dimensions other than relevance, for example popularity, review score, price, or recency. This is notably different from traditional information oriented search, including web search, where documents are surfaced in relevance order.

\section{Proceedings}
We had a broad mix of invited talks, paper presentations, poster discussions, and a panel discussion with active contribution and participation from both industry and academia.  Three papers were presented orally, and there was a poster session where all accepted papers (including the three) were presented. The poster session allowed extended interaction between the workshop attendees and the authors.

\subsection{Invited Keynotes}
The first invited keynote was presented by Eugene Agichtein (Emory University). Agichtein's talk,  \emph{``Modeling Explicit and Implicit User Behavior for Finding Answers on the Web''} described how explicit interactions captured using lightweight instrumentation of both search- and landing pages can be converted to attention and satisfaction signals; and how these signals can be used to improve selection of relevant documents, passages, and answers.  He also discussed his initial work on adapting these ideas to new interaction modalities.

The second invited keynote was presented by Xin Luna Dong  (Amazon).  Dong's talk, \emph{``Building a Broad Knowledge Graph for Products''} described her efforts in building a broad product graph.  This graph starts shallow with core entities and relationships, and allows easy addition of verticals and relationships in a pay-as-you-go fashion. She described her efforts on knowledge extraction, linkage, and cleaning to significantly improve the coverage and quality of product knowledge. Finally, she presented her progress towards her moon-shot goals including: harvesting knowledge from the web, hands-off-the-wheel knowledge integration and cleaning, human-in-the-loop knowledge learning, and graph mining and graph-enhanced search.

\subsection{Invited Short Talks}
In the first invited short talk, \emph{``Perspectives on search for e-commerce''}, Omar Alonso (Microsoft) presented insights on eCommerce search including the difficulty of defining relevance, metrics for measuring quality, and even the duration of a search session. The data often needs cleaning, and the line between brands and products is not clear. 

David Carmel (Amazon) presented the second invited short talk,\emph{``On the Relation between Products' Relevance and Customers' Satisfaction in Voice Shopping''}.  He discussed their observation that voice users sometimes purchase or engage with irrelevant search results. He presented an analysis and demonstrated its significance along with giving several hypotheses as to the reasons behind it (including customers' preferences, trendiness of the products, their relatedness to the query, the tolerance level of the customer, the query intent and the product price). 

The final invited short talk, \emph{``Getting people to describe fashion is hard''}, was presented by Owen Phelan (Zalando).  He gave an overview of how Zalando works with product data. He focused on the challenges they face in describing fashion products, and how these can have impact across the entire company and customer experience.

\subsection{Papers and Posters}
The workshop received 38 submissions from both industry and academia, and accepted 24 submissions (63\%). The submissions were reviewed by an international program committee of experts in the field formed from representatives of several eCommerce companies and academic institutions.  Each submission was reviewed by at least three reviewers. Of these 24 accepted submissions, three papers were chosen for oral presented, but all were presented in a poster session.

The first paper presented was by Momma et al., \emph{``Multi-objective Relevance Ranking''}~\cite{MichinariMomma}.
They introduce an augmented Lagrangian based method in a search relevance ranking algorithm to incorporate the multi-dimensional nature of relevance and business constraints, both of which are requirements for building relevance ranking models in production.  Experimental results show that the method successfully builds models that achieve multi-objective criteria much more efficiently than existing methods.

The second paper was \emph{``Learning Embeddings for Product Size Recommendations''} by Dogani et al.~\cite{KallirroiDogani}.  
They address the problem of size recommendation in fashion eCommerce with the goal of improving customer experience and reducing financial and environmental costs from returned items. They propose a novel size recommendation system that learns a latent space for product sizes using only past purchases and brand information.

The third paper, \emph{``Leverage Implicit Feedback for Context-aware Product Search''}, was by Bi et al.\cite{KepingBi}.
They leverage clicks within a query session as implicit feedback to represent users' hidden intents, which act as the basis for re-ranking subsequent result pages for a query. They also propose an end-to-end context-aware embedding model which can capture long-term and short-term context dependencies. Their results show that short-term context leads to much better performance compared with long-term or no context.

The last session of the workshop was a poster session with all the accepted paper submissions \cite{MichinariMomma, KallirroiDogani, KepingBi, eCOM1, eCOM2, eCOM3, eCOM4, eCOM5, eCOM6,eCOM7, eCOM8, eCOM9, eCOM10, eCOM11, eCOM12, eCOM13, eCOM14, eCOM15, eCOM16, eCOM17, eCOM18, eCOM19, eCOM20, eCOM21}. This session facilitated one-on-one interaction between the attendees and the authors.  It was well attended.

\subsection{Panel Discussion}
The topic of the panel was:~\emph{``eCommerce Discovery vs Web Search: Same or Different?''}.  The purpose of the panel discussion was to highlight how eCommerce and Web search are different, with the intention of drawing research interest into those areas.  The panel discussion was moderated by Vanessa Murdock (Amazon). The panelists were Estelle Afshar (The Home Depot), David Carmel (Amazon), Charles Clarke (University of Waterloo), and Xin Luna Dong (Amazon).

The moderator started the discussion by asking the panelists ``What web technology is most useful for eCommerce?''.  The answers were varied and included spelling checkers, query suggestion, query refinement, learning to rank, knowledge graphs, and leveraging user feedback.  

Differences were also suggested.  eCommerce user behavior and feedback is not as well understood as it is in web.  User emotion (happy or sad) is not apparent with voice systems such as Alexa.  In web search there is no dollar value to a search but with eCommerce there clearly is.

This lead to the next question -- the unique differences between Web search and eCommerce search.  It was generally agreed that although they are not entirely the same, they are also not entirely different.  The difficulty lies in understanding those differences (the objective function) and tailoring to that.  In the case of eCommerce search, this includes the dollar spending of the users.  In eCommerce search the long query tail is longer and sparse.  Optimising for long term customer satisfaction as opposed to short term gain was also discussed.

When asked what would be most useful for academics, the usual answers were given: data, money, review data, and Q\&A data.  It was agreed that industry and academia would benefit from working together.  Some data sets are already available, but benchmarks are needed.  From these benchmarks we can work out how to adapt the general IR frameworks to this specialised field.

Discussing scalability, the panelists highlighted the importance of knowing where your data comes from and what it is.  They also observed that there are people who specialise in scalability who will solve those problems once any solution exists.

Metrics and relevance were the next topic.  It was agreed that online and offline metrics complement each other and that relying on one over the other leads to a bad outcome.  Relevance was believed to be multi-objective, including: user satisfaction, loyalty, and so on.  Satisfaction is hard to measure because the user might be happy at the point of purchase, but unhappy three months later once the item fails.  A/B testing was discussed along with the difficulty of conducting such tests over the extended periods of time needed for eCommerce satisfaction determination.

Several other points were briefly discussed including: the similarity between search and recommendation, the value of precision of the first document (P@1) for mobile, the difference between ad search and eCommerce search, and different behaviours for known and unknown users.

Finally the panel was asked about open research challenges.  Afshar suggested the cold start problem for new items.  Dong is looking forward to a personal recommender that tells her, each morning, what she'll need to buy that day.  Carmel identified query understanding and the gap between queries and descriptions. Clarke identified the translation of industry problems into academic problems.

\section {High Accuracy Recall Challenge}
Jon Degenhardt (eBay) presented the details of the High Accuracy Recall Challenge, which was organised by the eBay Search Team but run as part of the workshop.

Non-relevance based results list sort orders (also known as deterministic sort orders) are common on eCommerce platforms.  This includes sorting a results list on price, distance, review score, and so on.  Web sort ordering is normally on relevance with no option of any other feature (such as date of publication).  There has been little or no formal study of search under such conditions.  The challenge was constructed to address both: methods of identifying relevant documents for deterministic sort, and ways of measuring the quality of the results.

eBay released a set of about 900,000 real-world listings from eBay's Collectibles category along with 150 popular queries from their query log (pertaining to the collection).  Each listing contained the title, the price, and the human readable category breadcrumb.  eBay also provided a set of binary relevance assessments (relevant or not-relevant) for about 110,000 query item pairs.

The task was to determine which of the 900,000 documents were relevant to each query and which were not.  The sort order of a results list is not typically known at the lowest depths of the search engine, so the task of identifying relevant documents must be done independent of the rank ordering.  That is, determining the recall set is a binary classification problem (is this document relevant or not?).  As such, the submitted runs were the results of a classifier -- besides, if price ordering was desired then a simple sort gives the results list (and likewise for time orders).

The challenge was hosted on evalAI\footnote{https://evalai.cloudcv.org/web/challenges/challenge-page/361/overview} and ran in three phases.  The first phase, unsupervised, required run submission before any training data (assessments) was released (May 17 - June 2).  For the second phase (June 3 - July 2), 60\% of the assessments were released and performance was measured on a held-out set of assessments.  For the final phase (July 3 - July 18), evaluation was on a second held-out set.

Performance was measured using various metrics including global Precision, Recall, and F1, query averaged Precision, Recall, and F1, and a new ``Price NDCG''.

In total sixteen groups participated.  The final leader boards are on the evalAI web site and we do not reproduce them here, instead we give only the top scores.  The top score using global F1 for the unsupervised phase was from ``Gotta Recall ‘em All'' with a score of 0.807.  For the supervised phase the top global F1 score was from ``JediOrder'' with a score of 0.845.  For the final phase, the top query averaged F1 score of 0.773 was from ``DeepBlueAI'', while the top Price NDCG score was 0.691 from ``JediOrder''.  Papers outlining the approaches are currently being collected and will be published online in due course.  It is clear already that a variety of very different approaches were used.

We believe that the challenge was a success and hope to further experiment with the metrics that were used, and with new approaches to identify a high accuracy recall base for later deterministic sorting.

\section{Conclusions and Future Directions}

The SIGIR 2019 Workshop on eCommerce (ECOM19) had a rich and diverse set of contributions and discussions. SIGIR was an excellent venue for the workshop -- and we thank them for their assistance in running the workshop. This workshop has proven to be a much needed forum for bringing together practitioners from industry and academia working in the eCommerce domain. The need to have an active community to discuss problems and potential opportunities in the eCommerce domain came to the forefront in the group discussion and panel sessions. 

This is the third time this workshop has been run at SIGIR.  the number of submissions has increased each year, as has the number of participants.  Workshop attendees continue to be enthusiastic in their support and look forward to future ECOM workshops at SIGIR.

\section{Acknowledgements}
We thank Estelle Afshar, Grigor Aslanyan, Manojkumar Kannadasan, and Aritra Mandal for taking minutes during the workshop.
We thank the presenters and authors for their contributions -- and from those contribution we took the summaries of their work.  Thanks also to the panelists, and participants for their contribution to the workshop.

\end{document}